\documentclass[%
 aip,
 numerical,
 rsi,
 amsmath,amssymb,
 reprint,%
]{revtex4-1}

\usepackage{dcolumn}
\usepackage{bm}

\usepackage[utf8]{inputenc}
\usepackage[T1]{fontenc}
\usepackage{mathptmx}

\usepackage[margin=25.4mm]{geometry}
\usepackage[utf8]{inputenc}
\usepackage{wrapfig}
\usepackage{color}
\usepackage{soul}
\usepackage{fancyhdr}
\usepackage{amsmath}
\usepackage{amssymb}
\usepackage{booktabs}
\usepackage{setspace}
\usepackage{csquotes}
\usepackage{bm}
\usepackage{graphicx}
\usepackage[english]{babel}
\usepackage{xspace}

\newcommand\apjs{\textit{ApJS}}
\newcommand\aaps{\textit{A\&AS}}
\newcommand\mnras{\textit{MNRAS}}

\newcommand\superbit{\textsc{SuperBIT}\xspace}

\newcommand\degree{\mbox{$^\circ$}\xspace}%
\newcommand\arcmin{\mbox{$^\prime$}\xspace}%
\newcommand\arcsec{\mbox{$^{\prime\prime}$}\xspace}%

\newcommand\utias{\affiliation{University of Toronto Institute for Aerospace Studies (UTIAS), 4925 Dufferin Street, Toronto, ON, Canada}}
\newcommand\dunlap{\affiliation{Dunlap Institute for Astronomy and Astrophysics, University of Toronto, 50 St. George Street, Toronto, ON, Canada}}
\newcommand\princeton{\affiliation{Department of Physics, Princeton University, Jadwin Hall, Princeton, NJ, USA}}
\newcommand\durhamcfai{\affiliation{Centre for Advanced Instrumentation (CfAI), Durham University, South Road, Durham DH1 3LE, UK}}
\newcommand\stewardobs{\affiliation{Department of Astronomy/Steward Observatory, 933 North Cherry Avenue, Tucson, AZ 85721-0065, USA}}
\newcommand\uoftastro{\affiliation{Department of Astronomy, University of Toronto, 50 St. George Street, Toronto, ON, Canada}}
\newcommand\uoftphysics{\affiliation{Department of Physics, University of Toronto, 60 St. George Street, Toronto, ON, Canada}}
\newcommand\jpl{\affiliation{Jet Propulsion Laboratory (JPL), California Institute of Technology, 4800 Oak Grove Drive, Pasadena, CA, USA}}
\newcommand\durhamcfea{\affiliation{Centre for Extragalactic Astronomy, Department of Physics, Durham University, Durham DH1 3LE, UK}}
\newcommand\durhamifcc{\affiliation{Institute for Computational Cosmology, Durham University, South Road, Durham DH1 3LE, UK}}
\newcommand\oslo{\affiliation{Institute of Theoretical Astrophysics, University of Oslo, Blindern, Oslo, Norway}}
\newcommand\sheffield{\affiliation{Department of Physics and Astronomy, The University of Sheffield, Hounsfield Road, Sheffield S3 7RH, UK}}

\begin{document}

\title{Robust diffraction-limited NIR-to-NUV wide-field imaging from stratospheric balloon-borne platforms\\-- \superbit science telescope commissioning flight \& performance}

\author{L. Javier Romualdez}
\email{javierr@princeton.edu}
\princeton

\author{Steven J.\ Benton}
\princeton

\author{Anthony M.\ Brown}
\durhamcfai
\durhamcfea

\author{Paul Clark}
\durhamcfai

\author{Christopher J.\ Damaren}
\utias

\author{Tim Eifler}
\stewardobs

\author{Aurelien A.\ Fraisse}
\princeton

\author{Mathew N. Galloway}
\oslo

\author{Ajay Gill}
\uoftastro
\dunlap

\author{John W.\ Hartley}
\uoftphysics

\author{Bradley Holder}
\utias
\dunlap

\author{Eric M.\ Huff}
\jpl

\author{Mathilde Jauzac}
\durhamcfea
\durhamifcc

\author{William C.\ Jones}
\princeton

\author{David Lagattuta}
\durhamcfea

\author{Jason S.-Y. Leung}
\uoftastro
\dunlap

\author{Lun Li}
\princeton

\author{Thuy Vy T.\ Luu}
\princeton

\author{Richard J.\ Massey}
\durhamcfai
\durhamcfea
\durhamifcc

\author{Jacqueline McCleary}
\jpl

\author{James Mullaney}
\sheffield

\author{Johanna M. Nagy}
\dunlap

\author{C.\ Barth Netterfield}
\uoftastro
\dunlap
\uoftphysics

\author{Susan Redmond}
\princeton

\author{Jason D.\ Rhodes}
\jpl

\author{J\"urgen Schmoll}
\durhamcfai

\author{Mohamed M.\ Shaaban}
\dunlap
\uoftphysics

\author{Ellen Sirks}
\durhamifcc

\author{Sut-Ieng Tam}
\durhamcfea

\date{\today}

\begin{abstract}
At a fraction the total cost of an equivalent orbital mission, scientific balloon-borne platforms, operating above $99.7\%$ of the Earth’s atmosphere, offer attractive, competitive, and effective observational capabilities -- namely space-like resolution, transmission, and backgrounds --  that are well suited for modern astronomy and cosmology.
\superbit is a diffraction-limited, wide-field, 0.5\,m telescope capable of exploiting these observing conditions in order to provide exquisite imaging throughout the near-IR to near-UV.
It utilizes a robust active stabilization system that has consistently demonstrated a $1\sigma$ sky-fixed pointing stability at 48\,milliarcseconds over multiple 1\,hour observations at float.
This is achieved by actively tracking compound pendulations via a three-axis gimballed platform, which provides sky-fixed telescope stability at $<$\,500\,milliarcseconds and corrects for field rotation, while employing high-bandwidth tip/tilt optics to remove residual disturbances across the science imaging focal plane.
\superbit's performance during the 2019 commissioning flight benefited from a customized high-fidelity science-capable telescope designed with exceptional thermo- and opto-mechanical stability as well as tightly constrained static and dynamic coupling between high-rate sensors and telescope optics.
At the currently demonstrated level of flight performance, \superbit capabilities now surpass the science requirements for a wide variety of experiments in cosmology, astrophysics and stellar dynamics. 
\end{abstract}

\maketitle

\section{Introduction}
\label{s:intro}

This paper presents the sub-arcsecond pointing and 50\,milliarcsecond image stabilization capabilities of the Super-pressure Balloon-borne Imaging Telescope (\superbit), for diffraction-limited, wide-field near-infrared (NIR) to near-ultraviolet (NUV)  imaging from a stratospheric balloon.
This first section introduces the science objectives that motivate these imaging capabilities, with a high-level description of the system architecture from the perspective of mechanical, optical, and control systems engineering.
Section~\ref{s:gondola} presents \superbit's best achieved performance to date, from the 2019 telescope commissioning flight; Section~\ref{s:analysis_and_discussion} analyzes the key technical improvements that enabled this performance, learned through earlier engineering test flights; and Section~\ref{s:forecast} predicts how the as-built performance could influence observing strategy during \superbit's upcoming long duration flight.
Detailed science forecasts, based on as-built performance, are being prepared for an accompanying {\em forecasting paper}.

\subsection{Scientific Applications}
The \superbit experiment is a balloon-borne telescope designed to provide diffraction-limited imaging over a 25\arcmin by 17\arcmin field-of-view (approximately 36 times larger than the {\it Hubble Space Telescope}'s Advanced Camera for Surveys) with an on-sky resolution of $<$\,0.3$\arcsec$. 
The platform utilizes the super-pressure balloon capabilities provided by the National Aeronautics and Space Administration (NASA), which offers mid-latitude long duration balloon (LDB) flights from 30 to 50+ days.
A telescope at 36\,km altitude is above 99.7\% of the Earth's atmosphere\cite{usstandardatmosphere}, enabling: 1) potentially diffraction-limited observations, with negligible atmospheric ``seeing'' $<10$\,milliarcseconds, and 2) space-like backgrounds and transmission throughout the wavelength range from near-ultraviolet (300\,nm) to the near-infrared (1000\,nm).

Within this wavelength range, the projected resolution and depth of \superbit imaging is sufficient to measure the (weak) gravitationally lensed shapes of distant (redshift $z\approx1$) galaxies behind foreground ($z\approx0.3$) clusters of galaxies.\cite{MasseyLensing}
Furthermore, \superbit's wide field-of-view allows an entire cluster to be imaged in a single pointing, including its connection to surrounding large-scale structure.
With imaging available in six selectable bands from 300 to 1000 nm, UV/blue photometry -- which is effectively inaccessible from ground-based instruments -- is particularly valuable for photometric redshift calibration, where cluster member galaxies can be identified via their 4000\,A break or the 3700\,A Balmer break in cluster dwarf galaxies for which this is suppressed.\cite{MasseyLensing, JonesSuperBIT}

For multiple observations of 100--150 clusters over a single super-pressure balloon flight, the high-quality cluster weak lensing masses estimated with \superbit would allow for the computation of fundamental cosmological parameters such as $\sigma_8$ and $\Omega_m$ at the level of experiments including Weighing the Giants and SPT-SZ\cite{2015ApJ...799..214B, 2015MNRAS.446.2205M}. 
In combination with X-ray or Sunyaev-Zoldovich (SZ) measurements, \superbit weak lensing maps of actively merging clusters would also be valuable for dark matter studies or calibration of cluster--mass observable relations. 
Additionally, \superbit's diffraction-limited imaging can mitigate de-blending calibration of ground-based cosmological surveys like the Large Synoptic Survey Telescope (LSST) \cite{2017ApJS..233...21R}, reducing that particular source of systematic uncertainty and leading to tighter constraints on cosmological parameters \cite{2018arXiv180706209P}.

Given the ability for balloon-borne platforms like \superbit to readily access, quickly implement, and flight verify cutting-edge technologies, high impact science goals can be realized at a fraction the economic and development time cost typical of equivalent space-borne implementations, with expected survey efficiencies rivaling similar ground-based applications.
In addition to cluster cosmology, some examples of prospective science goals enabled by \superbit include probes of dark matter sub-structure, strong gravitational lensing constraints on the Hubble constant, studies of galaxies' morphological evolution, UV-bright stars, and exo-planetary atmospheres.

\subsection{\superbit Architecture}
\label{ss:superbit_architecture}

The following is a brief description of the \superbit instrument flown on the September 2019 science telescope commissioning flight, the last
of a series of engineering test flights in advance of a long duration science mission.
Detailed descriptions of the mechanical \cite{StevenThesis, LunLi2016}, thermal \cite{RedmondThesis, Redmond18}, control systems \cite{RomualdezThesis18, Romualdez16, Romualdez18}, and software \cite{RomualdezThesis18} architectures for the 2019 and previous test flight configurations are available in the literature.

\subsubsection{Pointing \& Instrument Stabilization}

\begin{figure}
\centering
\includegraphics[width=0.5\textwidth]{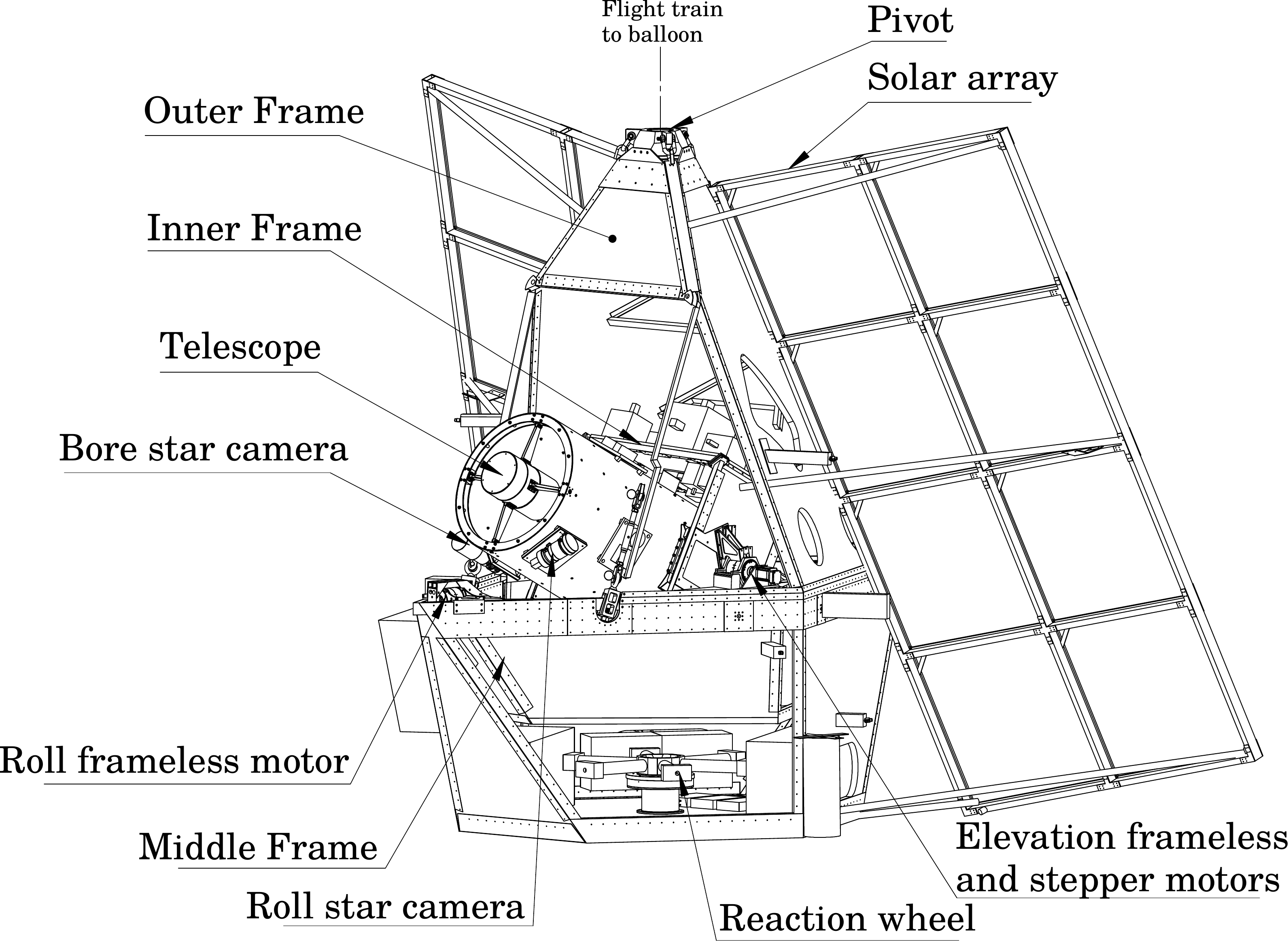}
\caption{\superbit 2019 gondola schematic with primary stabilization components identified; total mass at the pivot is 800\,kg, which includes CNES flight electronics; the \superbit 2019 gondola has 1600\,W solar power generation and 432\,Ah power storage systems.}
\label{fig:superbitoverview}
\end{figure}
\begin{figure*}
\centering
\includegraphics[width=1.0\textwidth]{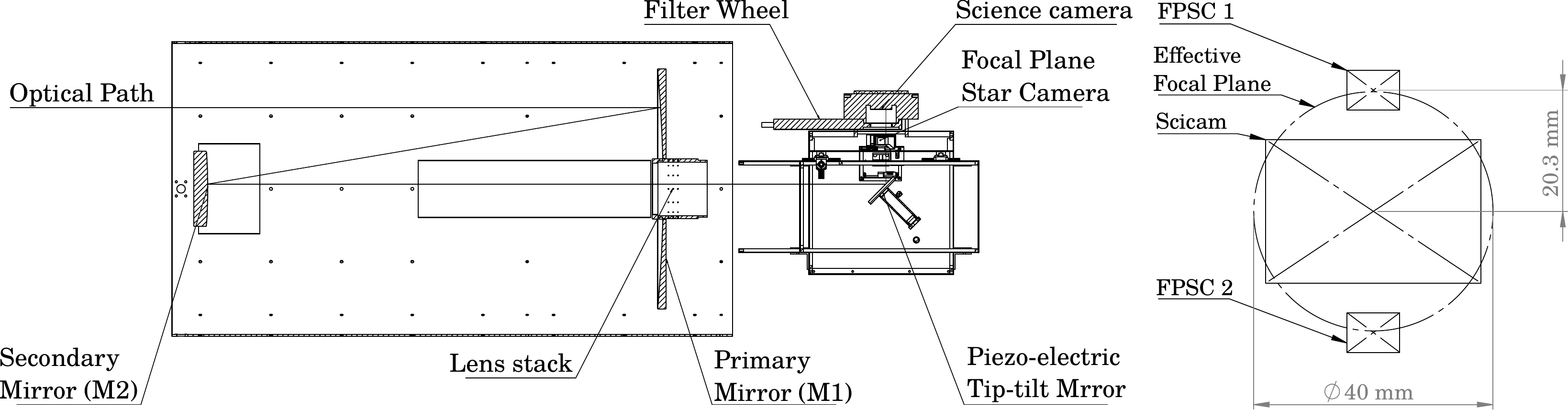}
\caption{(Left) cross-sectional view of \superbit science telescope commissioned during the 2019; the primary structure is comprised of a solid carbon fiber monocoque configuration with Invar components for thermally-sensitive components such as the secondary mount; the telescope is mounted in the \superbit inner frame on either side via opposing dovetail plates with compliance in the radial, axial, and cross-pitch directions to mitigate differential thermal contractions between the telescope and the aluminum frame; (right) the telescope focal plane layout including focal plane star cameras (FPSCs) and the science focal plane (scicam) with respect to the 40\,mm diameter \textit{effective} science-quality focal plane; the \textit{usable} or \textit{trackable} focal plane has a diameter of 55\,mm, which includes a region of reduced beam quality only sufficient for tracking.}
\label{fig:telescope}
\end{figure*}

From a purely engineering perspective, the \superbit platform, shown in Figure~\ref{fig:superbitoverview}, is a three-axis telescope stabilizer that is designed to provide sub-arcsecond stability for the science payload, namely the 0.5\,m NIR-to-NUV telescope, the scientific charge-coupled device (CCD) readout electronics, and accompanying back-end stabilization optics.
A series of three gimballed frames provide sub-arcsecond stabilization and control, which together correct both for gravity-driven compound pendulations induced by the balloon and flight train as well as the bulk sky rotation for long exposures (300--600\,s) over the science payload field-of-view ($\sim$\,0.5$^\circ$).
Gimbal roll and pitch control is facilitated per axis by frameless motors, each axially supported by flexure bearings that provide motion free from static friction, while a high-inertia reaction wheel facilitates yaw control and pendulation stability, with excess momentum dumped through the flight train to the balloon via the pivot connection \cite{RomualdezThesis18}.

Mounted to the science payload inner frame are two wide-angle (2--3$^\circ$) star tracking cameras -- one along the telescope boresight axis and the other orthogonal to it -- that provide absolute sky-fixed pointing references at 1--50\,Hz, while 1\,kHz rate gyroscopes (\textit{KVH Industries Inc.}) provide inertial stabilization feedback.
Altogether, science targets acquired with sub-arcminute-level accuracy are available with full three-axis sub-arcsecond stability for 30--60\,minutes per target, only limited mechanically by roll and pitch gimbal throw ($\pm 6^\circ$) and the full telescope pitch range (20--60$^\circ$).
This level of sky-fixed, three-axis stability is distinct from other balloon-borne stabilizers that provide only two-axis inertial stability\cite{WASP}.

\subsubsection{Telescope Optics \& Image Stabilization}
The \superbit science telescope (Figure \ref{fig:telescope}) is a modified-Dall-Kirkham $f/11$ design with optics sensitive from 300--1000\,nm 
(for a description of the pre-2019 engineering telescope, see 2015--2016 instrument papers\cite{Romualdez16,StevenThesis}).
As a custom-designed telescope (\textit{Officina Stellare}), the 0.5\,m conical primary, carbon-fiber monocoque body, and Invar components -- namely the secondary and lens stack mounting assemblies -- mitigate potential thermal gradients across optical components of the telescope assembly as well as variable mechanical loading due to elevation maneuvers.

Additionally, three equilaterally-placed linear actuators allow for the secondary mirror tip, tilt, and focus to be adjusted during operations to correct for changes in alignment and primary focus after launch or from variations in the bulk temperature profile of the telescope assembly due to diurnal cycles.
To correct for aberrations due to a spherical secondary mirror, a lens set near the back-end of the telescope assembly provides diffraction-limited imaging over a 55\,mm focal plane with a 37.5$\arcsec$/mm plate scale ($\sim$\,0.5$^\circ$ usable field-of-view).
Optics are thermally regulated through the telescope baffle to maintain a constant temperature profile and to mitigate large gradients when transitioning from day to night operations in the stratosphere.

In order to provide further image stabilization at the science CCD, a piezo-electric tip-tilt actuated fold mirror provides high-bandwidth (50--60\,Hz) focal plane corrections, which attenuates residual pointing jitter from the telescope stabilization systems down to 50\,milliarcseconds ($1\sigma$), well within the $<$\,0.3$\arcsec$ optical diffraction limit (more details in Section \ref{s:analysis_and_discussion}).
Sky-fixed feedback for the tip-tilt actuator is provided by a pair of focal plane tracking star cameras (FPSCs) -- one on either side of the science CCD -- while low noise rate gyroscopes (\textit{Emcore Corporation}) at 350\,Hz mounted directly to the telescope structure provide inertial feedback while actively correcting for the bulk latency and the limited bandwidth of the FPSCs.

\section{Instrument Performance}
\label{s:gondola}
\begin{figure}
\centering
\includegraphics[width=0.5\textwidth]{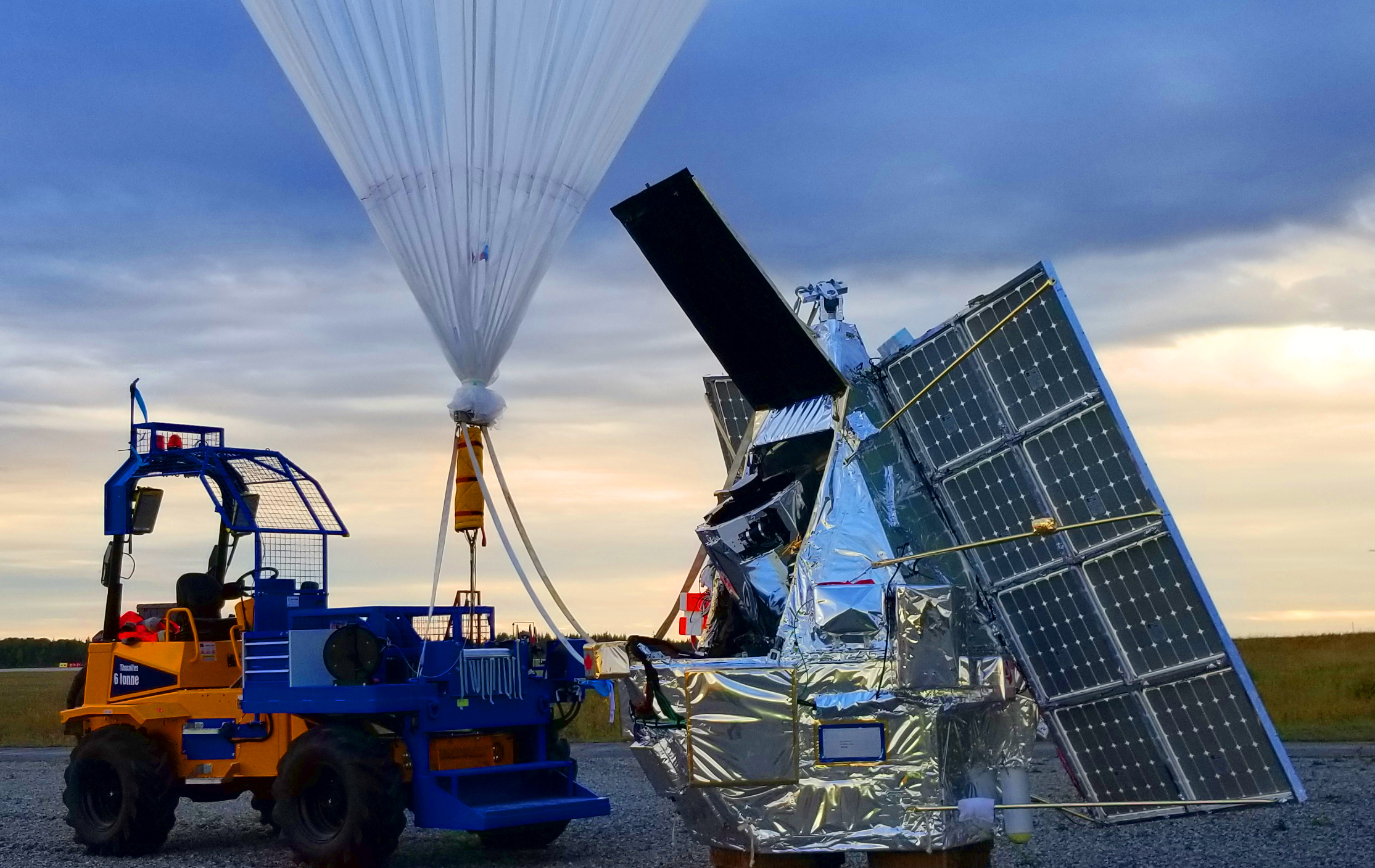}
\includegraphics[width=0.5\textwidth]{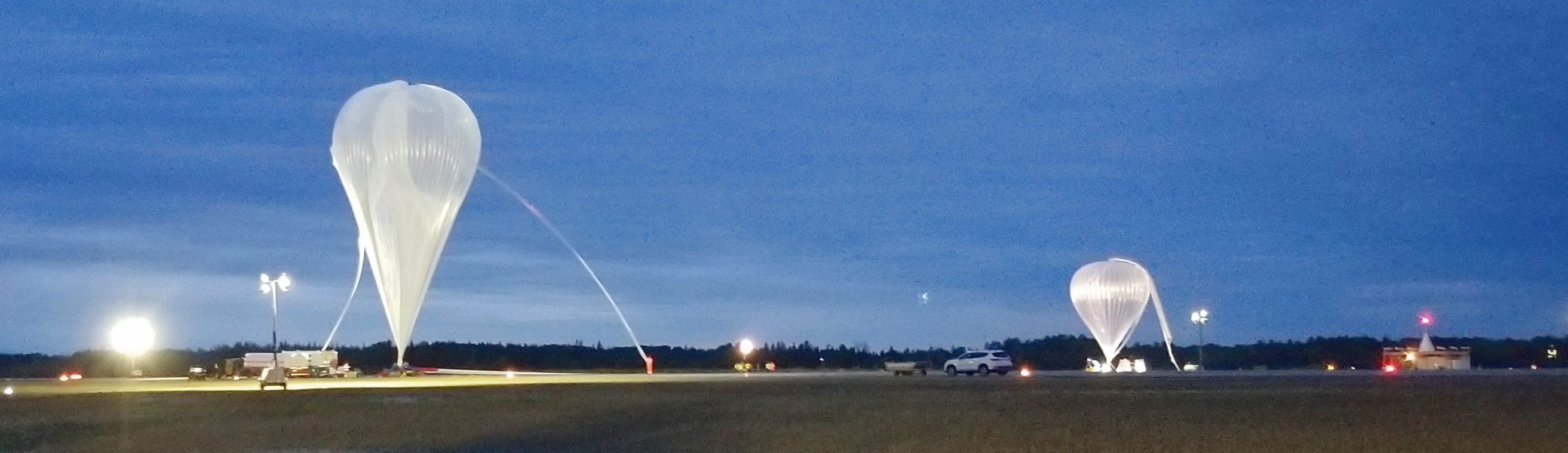}
\caption{\superbit 2019 pre-launch (site: Timmins, Ontario) secured by a launch support vehicle beneath the tow balloon (top); this smaller tow balloon (bottom, right) provides neutral buoyancy for launch and is secured during primary balloon (bottom, left) inflation.}
\label{fig:superbit2019_launch}
\end{figure}
The 2019 \superbit science telescope commissioning launch took place on September 17, 2019 at 20:34 GMT-4 with the \textit{Centre National d'\'Etudes Spatiales} (CNES) through the Canadian Space Agency (CSA) from the Timmins, Ontario launch site.
After a 2\,hour ascent to a minimum float altitude of 27\,km (89\,kft), the \superbit instrument was calibrated and aligned during the first 4 hours of operations, which was followed by 3.5\,hours of science observations.
Flight termination occurred on September 18, 2019 at approximately 14:00 GMT-4 after $\sim$\,7\,hours of daytime operations.

This section reports the best pointing and image stabilization performance achieved to-date from \superbit 2019 flight, which includes current \superbit performance results for telescope pointing stabilization, target acquisition accuracy, and focal plane image stabilization throughout typical operations.
To highlight particular challenges involved with demonstrating the achieved level of performance, comparisons with pre-2019 engineering test flights are provided (additional detail on pre-2019 performance is available in literature. \cite{Romualdez16, Romualdez18}).

\begin{figure*}
\centering
\includegraphics[width=1.0\textwidth]{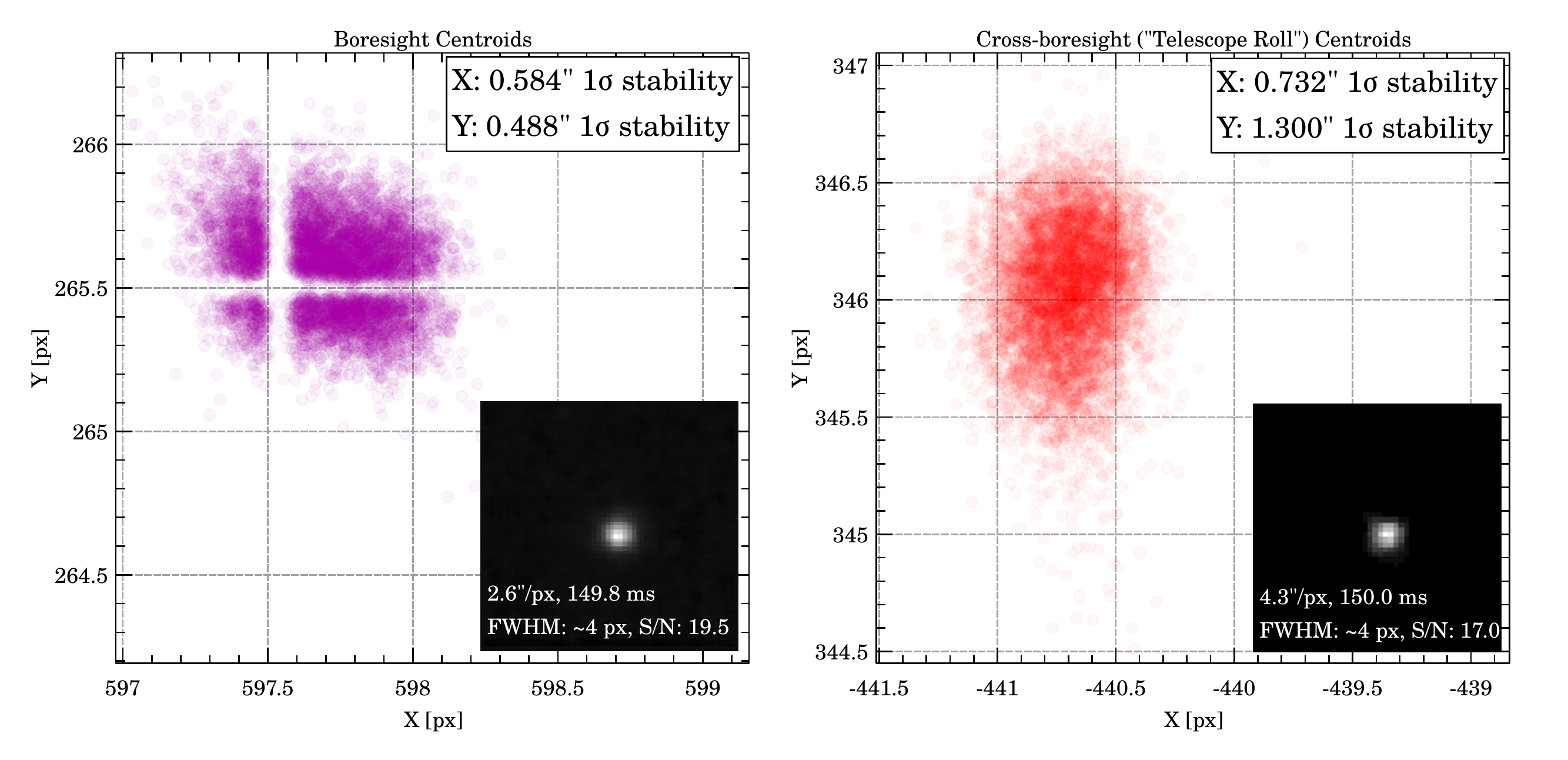}
\caption{Telescope pointing stability as measured by the boresight (left) and cross-boresight (right) star cameras during a typical long timescale telescope tracking run ($> 30$\,minutes) from the \superbit 2019 flight, where each data point is a single sky-fixed star centroid; typical star camera centroid thumbnails are shown, where boresight and cross-boresight sky-equivalent pixel scales are 2.6$\arcsec$ and 4.3$\arcsec$, respectively; sub-pixel structure (i.e. gaps between pixels) are clearly resolved at pixel boundaries (with a 0.5\,px offset) due to non-linear pixel response.
}
\label{fig:telescope_stability}
\end{figure*}
\begin{figure*}
\centering
\includegraphics[width=1.0\textwidth]{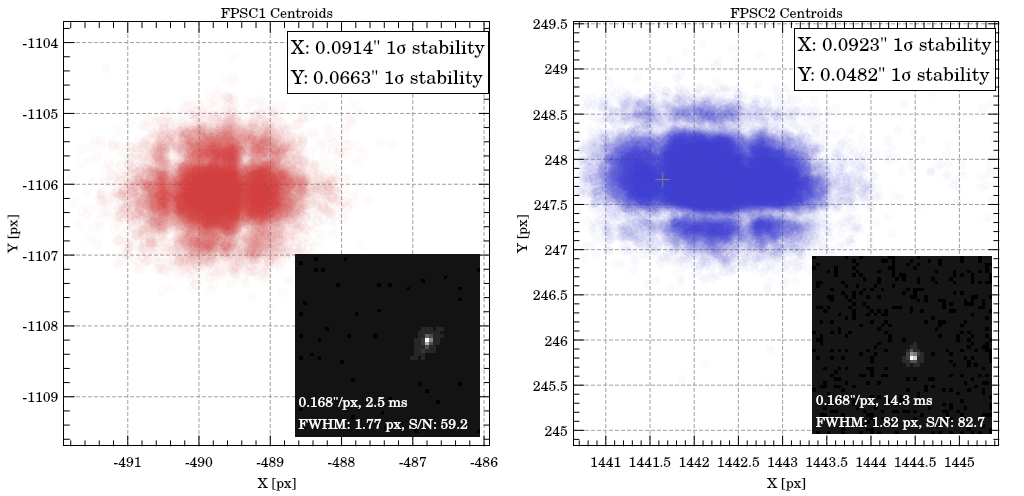}
\caption{Actively stabilized FGS tracking stability as measured by each of the focal plane star tracking cameras (FPSCs), where each data point is a single sky-fixed star centroid; one representative 6\,minute tracking run on FPSC1 (left) and FGS tracking on FPSC2 (right) during the telescope stabilization run in Figure \ref{fig:telescope_stability} are shown; typical star camera centroid thumbnails are shown, where the sky-equivalent pixel scale for each FPSC is 0.168$\arcsec$; sub-pixel structure (i.e. gaps between pixels) are clearly resolved at pixel boundaries (with a 0.5\,px offset) due to non-linear pixel response.}
\label{fig:image_stability}
\end{figure*}

\subsection{Telescope Stabilization}
As described in Subsection \ref{ss:superbit_architecture}, the \superbit telescope itself is stabilized about all three rotational degrees-of-freedom via a series of gimballed frames with inertial feedback from rate gyroscopes and sky-fixed reference feedback from star tracking cameras.
When a target is requested in right ascension (RA) and declination (Dec), each of the gimbals slew in unison to the calculated target gimbal angles in order to point the telescope to the corresponding azimuth (Az) and elevation (El) coordinates on the sky whilst offsetting the middle and inner frames to maximize tracking time on the sky with fixed field rotation (FR). 
Once a lost-in-space solution is acquired at the target coordinates\cite{dustin}, the gimballed frames are iteratively honed towards the desired target until the absolute error on the sky is within a commandable star camera subframe threshold, which for \superbit is typically $0.5\arcmin$.
Following this, star camera centroids are then used for higher rate feedback, which provides absolute, sky-fixed stability while concurrently correcting for biases in the rate gyroscopes that provide inertial stability.

During the 2019 flight, \superbit tracked and stabilized over several telescope tracking runs for a number of alignment calibration and potential science targets of interest.
For a typical long timescale telescope stabilization run ($\sim$0.5\,hr), Figure \ref{fig:telescope_stability} shows centroid distribution plots in the boresight and cross-boresight (or ``telescope roll'') star cameras as well as a representative point-spread-function (PSF) per tracking camera. 
Analyses and discussions on this level of sub-arcsecond pointing performance are provided in subsection \ref{ss:stabilization}.

\subsection{Image Stabilization}
Once telescope stabilization is established, a high-bandwidth, piezo-electrically actuated, tip-tilt fold mirror further stabilizes the focal plane. 
As described in subsection \ref{ss:superbit_architecture}, sky-fixed feedback is obtained from a pair of focal-plane tracking star cameras, each equally separated about the center of the science camera CCD by 29.3\,arcminutes (see Figure \ref{fig:telescope}).  
In addition to providing inertial feedback, a set of low noise rate gyroscopes directly and rigidly coupled to the telescope frame itself increases the effective bandwidth and corrects the bulk image processing latencies inherent to each of the focal plane star cameras. 
Altogether, this tracking system comprises the fine guidance system or FGS.

To be able to effectively track the residual perturbations on the focal plane, the expected peak-to-peak variation from the telescope stabilization stage must be well within the maximum throw of the tip-tilt stage. 
With knowledge of the 2019 back-end optics geometry, notably the distance between the tip-tilt stage and focal plane ($\ell = 169$\,mm) and the maximum stage-centered throw of the FGS ($\theta = 2$\,mrad), the maximum sky-equivalent pitch throw $\phi_{pitch}$ and cross-pitch throw $\phi_{xpitch}$ of the FGS are given by
\begin{eqnarray}
\phi_{\rm pitch} &=& p\cdot \ell \cdot \theta \ , \ \phi_{\rm xpitch} = \phi_{\rm pitch}\cdot \cos(45\deg)\\
&\Rightarrow & p = \tfrac{206264.8\arcsec}{D\times f}
\label{eq:fgs_throw}
\end{eqnarray}
where $p = 0.168\arcsec$/px is the FPSC plate scale, $D = 500$\,mm is the primary mirror diameter, and $f = 11$ is the $f$-number for the \superbit 2019 science telescope.
Due to the $90^\circ$ fold in the optical path (see Figure~\ref{fig:telescope}), the effective throw of the cross-pitch axis is reduced by a factor $\cos(45^\circ)$ compared to the pitch axis.
To highlight the performance shown previously, the $\phi_{\rm pitch}=12.61\arcsec$ and $\phi_{\rm xpitch}=8.92\arcsec$ requirements on stabilization are at least 15 times the demonstrated $1\sigma$ telescope stabilization.
Note however that this specification does not include potentially long timescale mechanical or thermal drifts between the boresight and focal plane star cameras since such variations are unknown at the telescope tracking level without feedback information from the tip-tilt stage.
These longer timescale effects are reported in the following subsection and further discussed in Section \ref{s:analysis_and_discussion}.

Once trackable stars were reliably and accurately obtained on one or both of the FPSCs during the first half of the 2019 flight, several dedicated image stabilization runs calibrate the FGS and to align telescope optics.
Figure \ref{fig:image_stability} demonstrates the typical image stabilization performance post-calibration by evaluating focal plane star camera centroid distributions as a sky-fixed metric for FGS corrections, in response to residual disturbances from the outer telescope stabilization loop.
Analyses and discussions on FGS performance, including FPSC tracking depth, FPSC beam size, and overall image stability are provided in subsections \ref{ss:stabilization} and \ref{ss:diffraction_limited}.

\subsection{Target Acquisition}
\label{ss:target_acquisition}
\begin{figure}[t]
\centering
\includegraphics[width=0.49\textwidth]{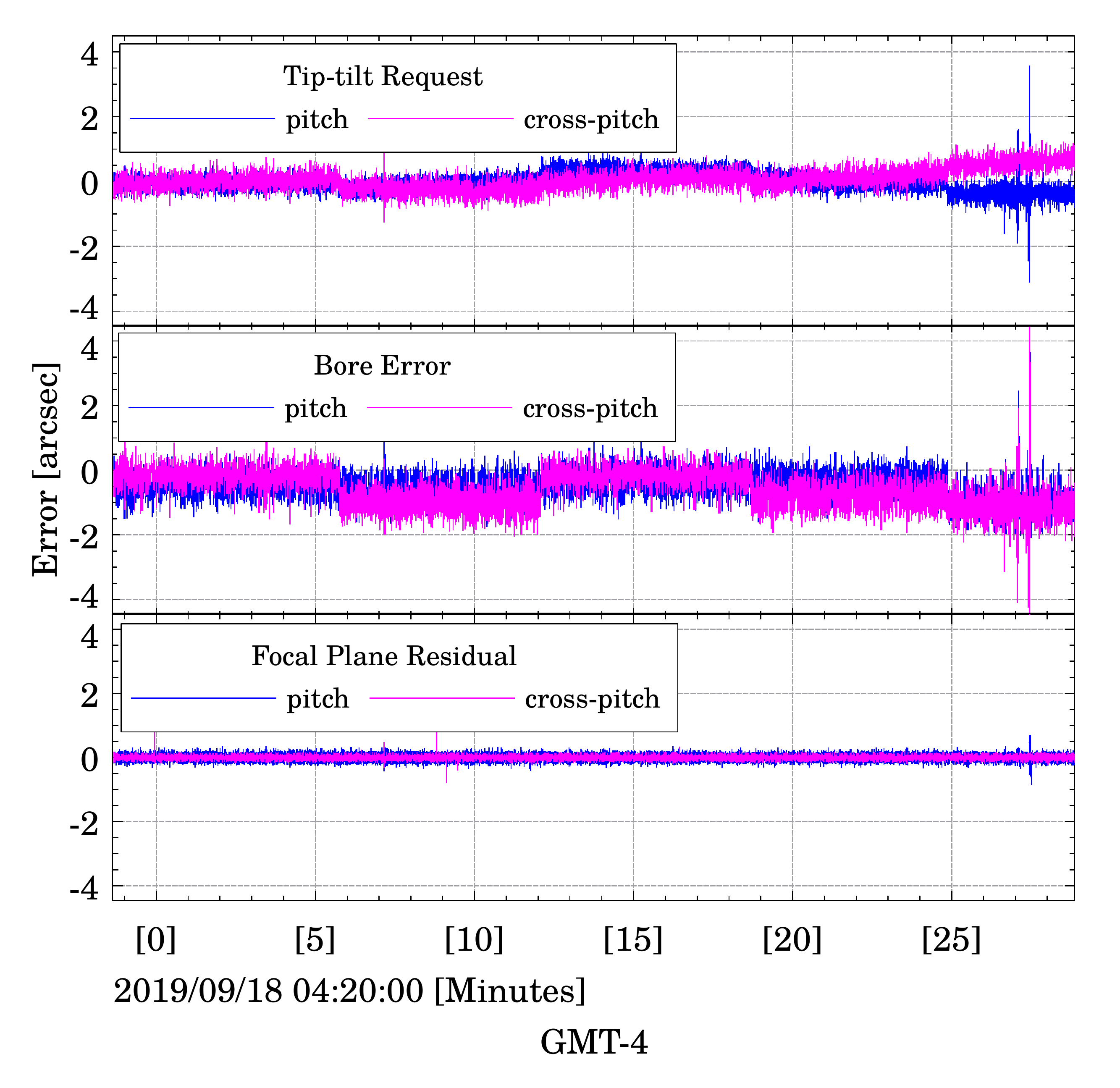}
\caption{(Top) absolute pointing residual plot illustrating the drift of the FGS tip-tilt mirror command during the image stabilizaiton run shown in Figure \ref{fig:image_stability}; this is not the result of telescope pointing drift, as would be seen by the bore star camera centroid residuals (middle), or telescope focal plane drift, as would be seen by the FPSC centroid residuals (bottom), which implies that this is likely relative pointing drift between that the boresight star camera and the telescope focal plane; despite this, the $<$1$\arcsec$ drift is maintained well within the $8.92\arcsec$ (or $\pm 4.46\arcsec$) cross-pitch throw of the FGS; on-sky directions for all residual plots are given in pitch and cross-pitch to distinguish from telescope El and cross-El; randomized commanded dithering steps can be seen at 5\,minute intervals (top \& middle) but are removed by the FGS at the telescope focal plane (bottom).}
\label{fig:pointing_drift}
\end{figure}
To find adequate guide stars for image stabilization, the \superbit stabilization platform must have sufficiently accurate absolute pointing to place a target within the roughly $5.5\arcmin\times 4\arcmin$ field-of-view of either focal plane star camera. 
Although the pointing and tracking systems are capable of acquiring targets to much higher precision (within the 0.26$\arcsec$ star camera centroiding resolution), \superbit target acquisition requires only arcminute-level repeatability, because of its wide field-of-view and absolute sky-fixed feedback.
If higher accuracy is required post-flight, absolute pointing information can be reconstructed from flight data and confirmed directly with astrometry\cite{dustin} on science camera images.
Over extended tracking runs, relative drift between the boresight and focal plane star tracking cameras is shown in Figure~\ref{fig:pointing_drift} to be minimal, which contrasts the large 10--20 arcsecond pointing drift observed during pre-2019 test flights.
This reflects both the thermo-mechanical and the opto-mechanical stability of the 2019 telescope optics, back-end optics, and star camera mounts, a detailed discussion of which is provided in subsection \ref{ss:opto_mech_stability}.
To quantify this effect, the sky-equivalent FGS command shown in Figure~\ref{fig:pointing_drift} measures how the focal plane would have moved with respect to the boresight star camera had the FGS not been actively and continuously correcting for structural or kinematic disturbances.
Plausible sources for residual discrepancies between optics and telescope pointing are discussed in \ref{ss:opto_mech_stability}.

\begin{figure}
\centering
\includegraphics[width=0.5\textwidth]{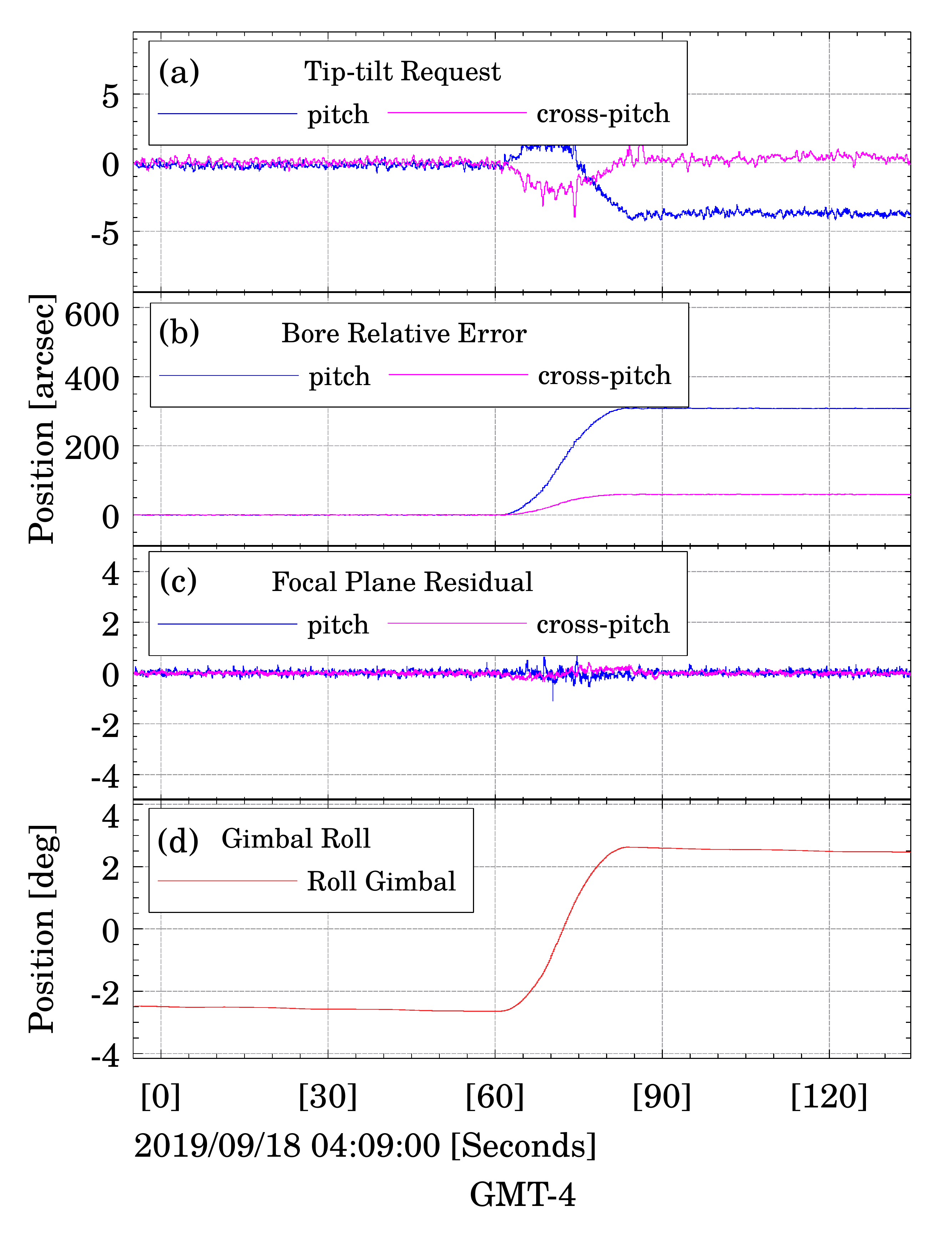}
\caption{Target re-acquisition and continuous target tracking during simultaneous roll and fine pitch gimbal resets; gimbal frames track in unison about the FPSC target star whilst the FGS attenuates perturbations at the 2--4$\arcsec$ level (a) down to $<$0.2$\arcsec$ (c) during the reset; the boresight star camera centroids (b) show a mean shift post-reset due to an offset pointing vector with the FPSC(s) as the roll gimbal is slewing (d); the maximum sky-equivalent throw of the FGS in pitch and cross-pitch is 12.61$\arcsec$ and 8.92$\arcsec$ respectively.}
\label{fig:gimbal_reset}
\end{figure}

For \superbit, target acquisition repeatability and target re-acquisition are also important factors contributing to overall survey efficiency during science operations.
In particular, a fixed-time overhead is required to reset the roll and pitch gimbals once either axis has reached its maximum usable throw, which is fundamentally limited by the flexure bearings ($\pm 6\degree$ smooth tracking range), as mentioned in subsection \ref{ss:superbit_architecture}.
To minimize the time required to recover image stabilization after a reset, all three gimballed axes are slewed in unison about the FPSC target star in a way that resets the roll and pitch gimbals while maintaining the target star within the FGS full throw.
Figure~\ref{fig:gimbal_reset} demonstrates the efficacy of this approach from a typical gimbal reset during the \superbit 2019 flight.
During this kind of reset, which takes place over about 20\,seconds, science camera exposures are temporarily halted, and the net effect on subsequent science camera exposures post-reset is a 1-6\,degree field rotation about the current tracking FPSC depending on the target. 

\section{Analysis \& Discussion}
\label{s:analysis_and_discussion}

\subsection{Telescope \& Image Stabilization}
\label{ss:stabilization}
As presented in the previous section, telescope stabilization performed exceptionally well, where the maximum throw of the FGS is 15 times the worst-case $1\sigma$ pitch/cross-pitch pointing stability over even the longest tracking timescales.
Even though telescope roll stability was only maintained at the arcesecond level, as shown in Figure \ref{fig:telescope_stability}, the sensitivity of the telescope focal plane to roll perturbations is significantly lower than in pitch and cross-pitch, where a $1\arcsec$ roll motion over a $0.5\deg$ angular separation between the telescope and the boresight star camera is $<$10\,milliarcseconds in the worst case.
Furthermore, the fidelity of both the telescope and image stabilization performance (in Figures \ref{fig:telescope_stability} and \ref{fig:image_stability}, respectively) is highlighted by the clear sub-pixel structure observed in the centroid distributions, where the finite size of the pixel gaps on the respective imaging sensors is clearly resolved near the pixel boundaries (with a 0.5\,px offset).

\subsubsection{Stabilization Trade-offs}
\begin{figure}
\centering
\includegraphics[width=0.5\textwidth]{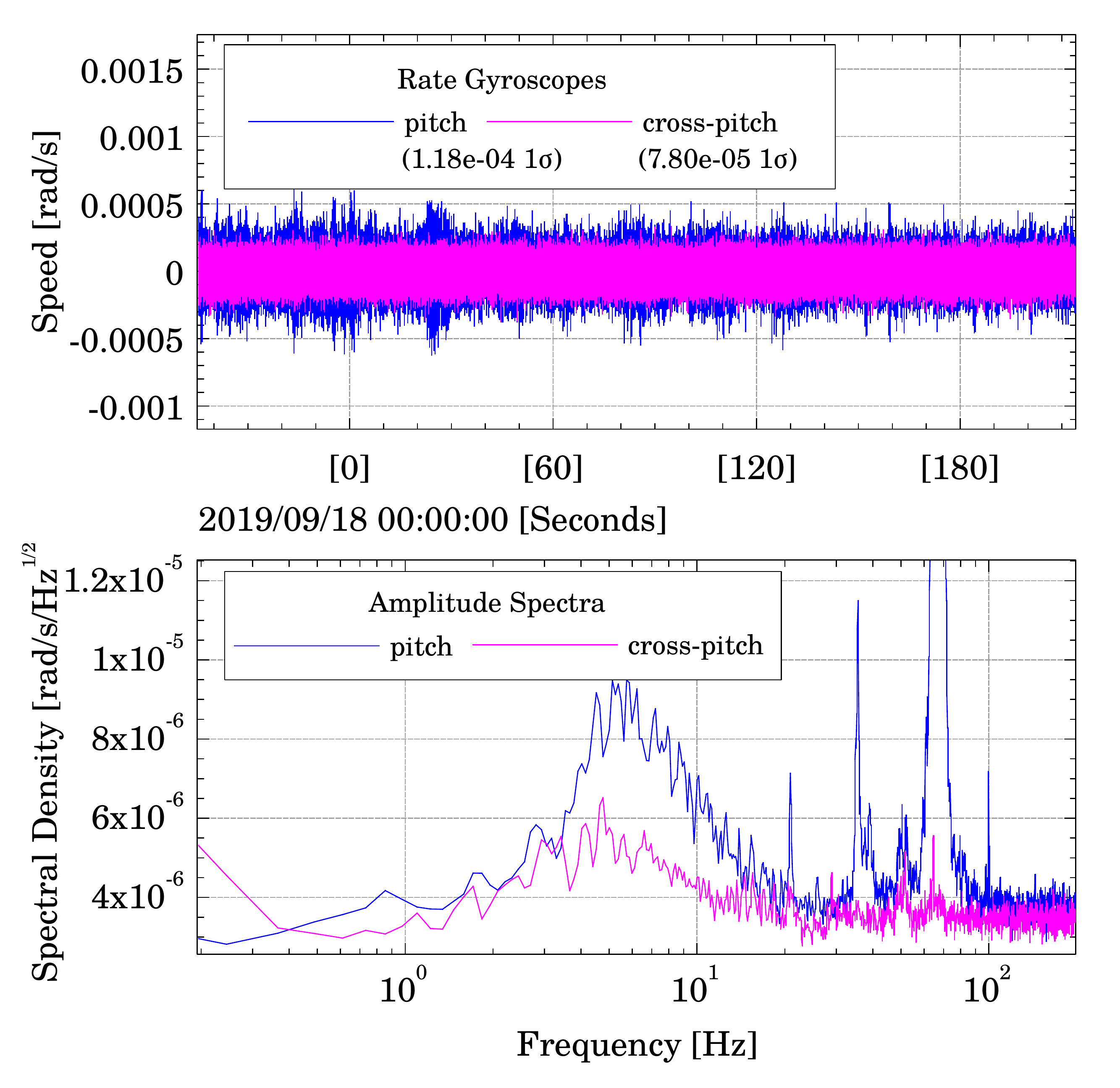}
\caption{Rate gyroscopes residuals after gain calibration of the telescope stabilization stage during the 2019 \superbit telescope commissioning flight; timestreams (top) and amplitude spectra (bottom) are given in the telescope pitch and cross-pitch axes during a period where the gondola was only being inertially stabilized (i.e. rate gyroscopes controlled to zero speed with no sky-fixed feedback); the $1\sigma$ noise measured in the cross-pitch axis is 35\% below the $1\sigma$ noise measure in pitch.}
\label{fig:pitch_overcontrol}
\end{figure}

The high performance of the \superbit telescope stabilization platform suggests that there may have been potential trade-offs between telescope stabilization and image stabilization that could have further improved the overall attenuation and fidelity of the FGS in correcting for the residual telescope disturbances.
Figure \ref{fig:pitch_overcontrol} shows the zero-speed inertial stabilization (i.e.\ rate gyroscopes only) immediately following the tuning phase of the 2019 flight.

\begin{figure*}
\centering
\includegraphics[width=1.0\textwidth]{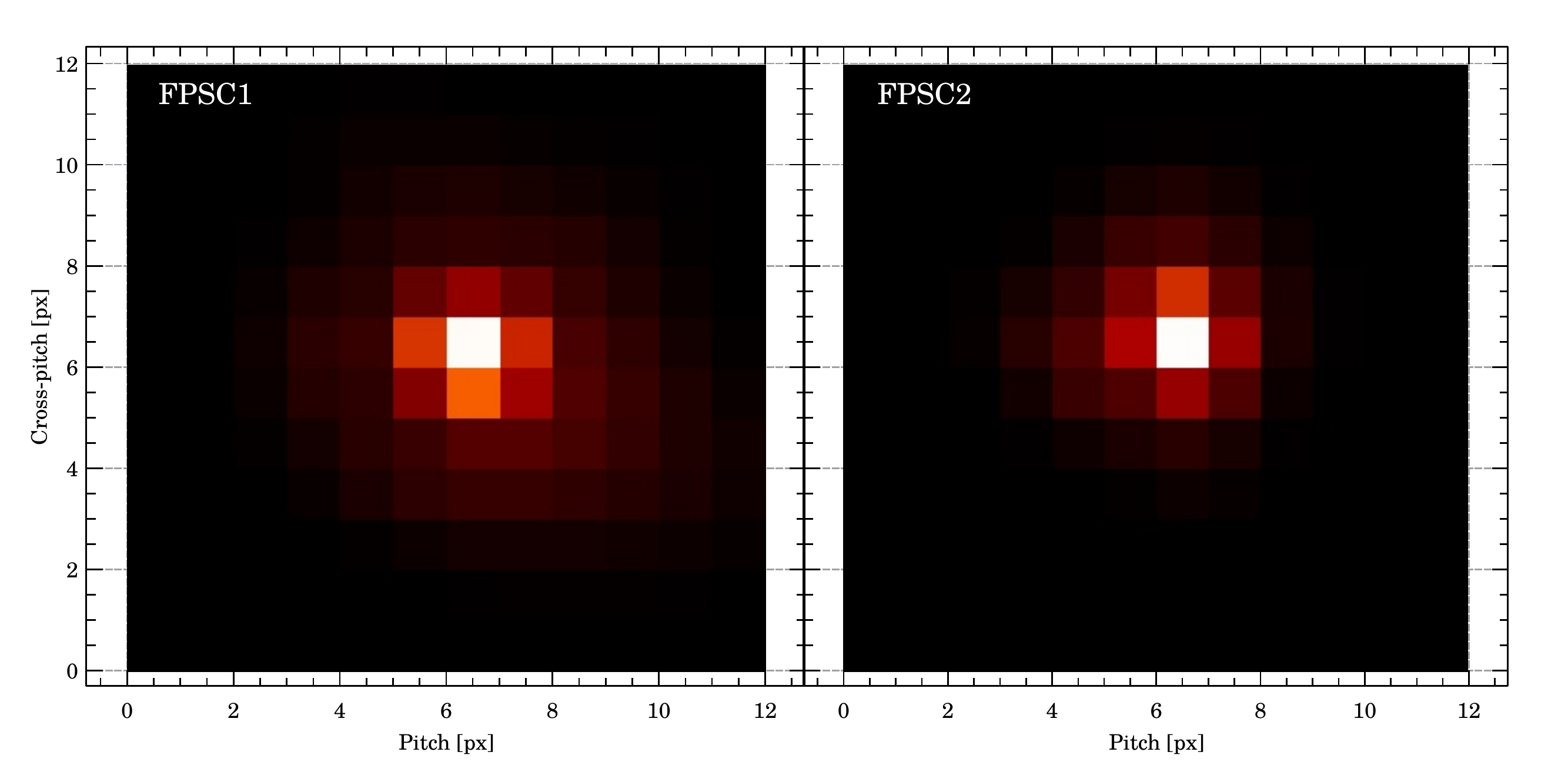}
\caption{Co-added FPSC PSFs integrated over typical science camera exposures during the image stabilization run from Figure \ref{fig:image_stability}, which represents the combination of pointing jitter as well as the effective optical beam through the \superbit science telescope optics; FPSC pixel scale is 0.168$\arcsec$/px for an measured co-added FWHM of $0.400\arcsec$ on FPSC1 and $0.350\arcsec$ on FPSC2; correcting for pointing jitter at a FWHM of 0.113--0.217$\arcsec$ yields an optical PSF FWHM of 0.273-0.302$\arcsec$, which agrees well with the simulated diffraction-limited PSF.}
\label{fig:fpsc_coadd}
\end{figure*}

Comparing pitch and cross-pitch control, it is clear that there is a noticeable level of over-control in the pitch axis, which is characterized by the wide-band feature in the pitch spectrum centered at $\sim$7\,Hz.
This idea of pitch over-control is supported by the fact that the pitch spectrum at low ($<$1\,Hz) frequencies has been pushed below the noise floor seen at higher frequencies. 
Although there is some over-control in the cross-pitch axis as well, the cross-pitch over-control is 42\% lower than in the pitch axis, which suggests that the pitch control gains may have been slightly over-tuned.
The overall result of this over-tuning is a level of high frequency rate gyroscope noise from telescope stabilization leaking into the image stabilization stage.

Looking at Figure \ref{fig:image_stability}, this effect is directly observed in the 48\% reduction in performance in the FGS pitch-oriented axis compared to cross-pitch.
It is therefore reasonable to assert that a reduction in pitch gains, though potentially reducing performance at the telescope stabilization stage at lower frequencies, would have reduced higher frequency rate gyroscope noise that inadvertently degraded image stabilization in pitch.

Overall, it is clear from the results that the 50\,milliarcsecond stability requirement was consistently achieved during flight in the cross-pitch, and despite the fact that there is indeed a clear asymmetry in FGS performance due to pitch over-control effects, the level to which the FGS performed is still sub-pixel on the science CCD and below the diffraction limit ($<$\,0.3$\arcsec$).
This level of performance met science requirements; in part due to time constraints, the 2019 performance was deemed adequate to begin science operations for the remainder of the 2019 flight. 
Additional tuning would have likely yielded improved performance in the pitch axis and, potentially, in image stabilization overall.

\subsection{Diffraction-Limited Performance}
\label{ss:diffraction_limited}
\subsubsection{FPSC Beam Size}

The image stabilization results in Figure \ref{fig:image_stability} provide preliminary insights into the optical performance of the 2019 \superbit science telescope, which not only influences the quality of the resulting science images, but is directly related to the limiting depth and fidelity of FGS centroiding and, therefore, focal plane tracking performance.
To quantify the sharpness of the star imaged on the FPSCs, the full-width half-max (FWHM -- in pixels, denoted $W$) reported in flight (and shown in Figure~\ref{fig:image_stability}) is estimated by
\begin{eqnarray}
\hat{W} = \frac{4\log{2}}{\pi}\sqrt{\frac{\sum_i p_i}{\mbox{max}_i p_i}}
\label{eq:fwhm_est}
\end{eqnarray}
where $\sum_i p_i$ is the sum of all the pixels and $\mbox{max}_i p_i$ is the maximum/peak pixel value in the background-subtracted image subframe.

Due to the small number of pixels that contribute to the sum post-background-subtraction, this estimator $\hat{W}$ is inherently subject to pixel noise and is potentially biased due to variation in background estimates.
To correct for this bias, simulated Gaussian sources were tuned such that $\hat{W}$ and background noise matched those observed and measured during flight.
Over 10000 simulations, the input FWHM ($W$) used to match the flight images were then used to generate a calibration curve given by $W \simeq 1.307\hat{W}-0.705$.
For the majority the 2019 flight, the estimated FWHM values on FPSC2 were within the $\hat{W} = $1.7--1.9\,px range per image, which correspond to bias-corrected FWHM values of $W =$1.52--1.78\,px or 0.255--0.300$\arcsec$ on the sky.
A representative measure of FWHM on FPSC1, which had considerably fewer image stabilization runs, is $\hat{W} = $1.75\,px corresponding to $W =$ 1.58 px or 0.265$\arcsec$ on the sky.

For a typical 300 second science camera exposure, Figure \ref{fig:fpsc_coadd} shows the co-added PSF on the FPSCs, which effectively captures the optical PSF convolved with the pixel response, and the measured pointing jitter during image stabilization.
Fitting a two dimensional Gaussian to the PSF yields a cumulative $0.400\arcsec$ and $0.350\arcsec$ FWHM on FPSC1 and FPSC2, respectively.
After deconvolving the pointing jitter with a FWHM from  0.113--0.217$\arcsec$ based on Figure \ref{fig:image_stability}, this implies an optical PSF with a 0.273--0.302$\arcsec$ FWHM.

\begin{figure}
\centering
\includegraphics[width=0.5\textwidth]{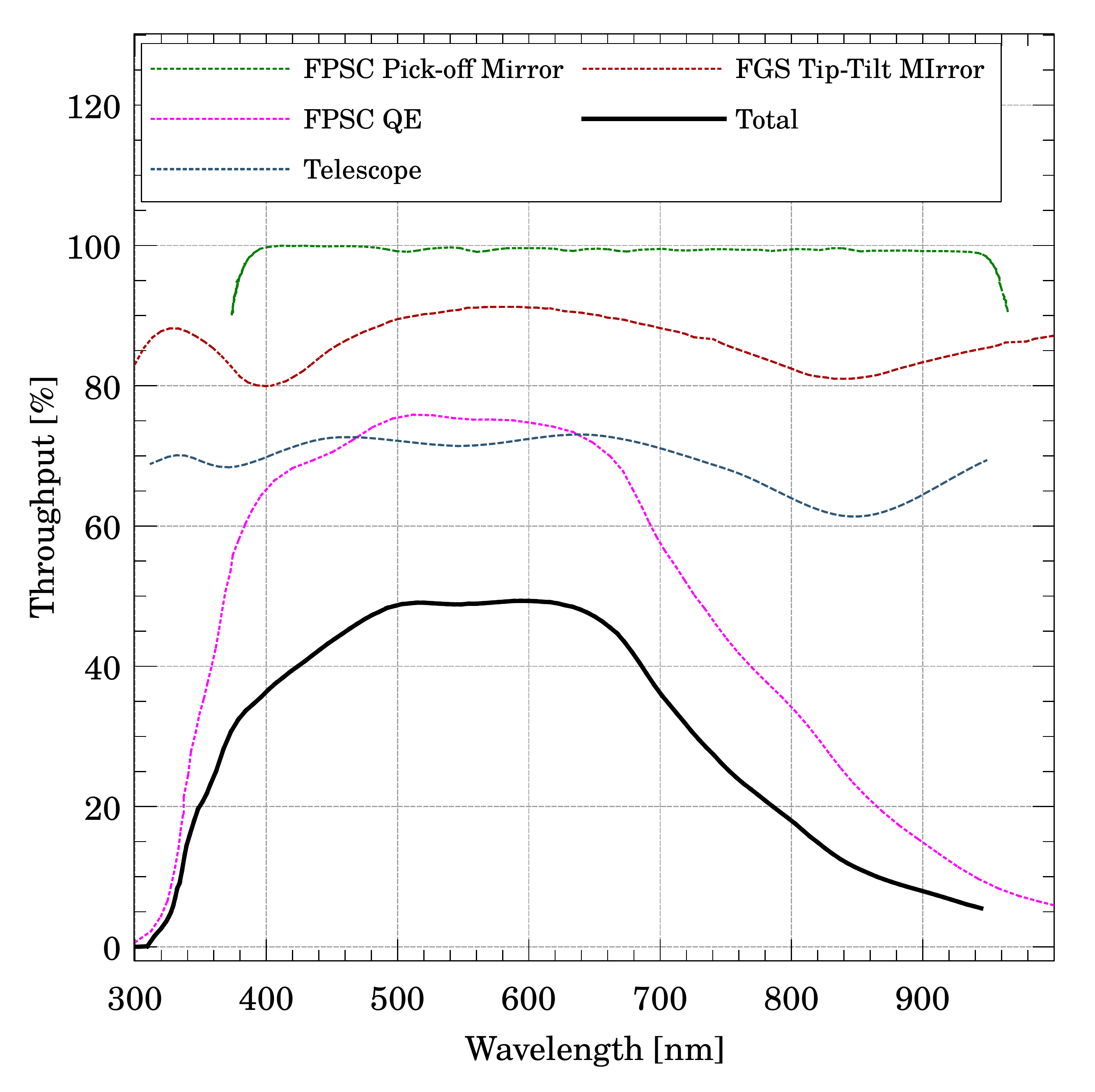}
\caption{Theoretical per-band optical efficiency at the FPSCs focal plane, which incorporates the following: optical characteristics, namely the transfer curves for the primary mirror, secondary mirror, and the lens stack combined; back-end optical characteristics, including the tip-tilt fold mirror and the FPSC pick-off mirror; and the nominal FPSC quantum efficiency response. For a flat spectrum source, the effective band center of the FPSCs is 600\,nm, with 89\% bandwidth and a band-average throughput of 38\%.}
\label{fig:fpsc_efficiency}
\end{figure}

The expected FPSC PSF can be estimated from simulation by taking into account the effective throughput at the FPSC (Figure \ref{fig:fpsc_efficiency}) and integrating the nominal simulated optical beam per-band.
From this, the theoretical pixel-convolved FWHM is $0.273\arcsec$ on the sky, which agrees with the measured value remarkably well compared to the best-case measured PSF on FPSC2 ($<$1\%) and within 10\% for the worst-case.
Note that, given its broadband spectral response, the FPSC PSF is markedly wide compared to the typical PSF expected in many of the \superbit science bands; however, the FPSCs provide higher PSF spatial resolution due to 20\% smaller pixels compared to the science camera.

As shown in Figure \ref{fig:fpsc_coadd}, the measured PSF in the FPSCs on opposite sides of the field of view differ at a level of 50\,milliarcseconds, near the demonstrated limit of tracking stability.  
There are a variety of potential causes for this PSF variation.  
For example, a degree of optical misalignment may have been present at the secondary, causing an asymmetry in the aberrations at the location of the two FPSCs, which are offset 20.3\,mm from the optical axis (Figure \ref{fig:telescope}).  
Pointing jitter over the FPSC exposure time is also convolved with the optical PSF per image, but the level to which this plays a role is limited by the short exposure times typical on the FPSCs (nominally $<$\,20\,ms).
Relative focus stages allow for FPSCs' focus positions to be calibrated independently, so the 15\% broader PSF observed in FPSC1 compared to FPSC2 could be attributed to relative defocus between the two tracking camera sensors.
All these can easily be improved through more regular calibration (e.g. daily) during a longer duration mid-latitude science flight.

\subsubsection{FPSC Tracking Depth}
\begin{table*}
\caption{\label{tab:fpsc_mags}Estimated tracking depth on the FPSCs based on uniquely identified guide stars from the \superbit 2019 flight; the tracking depth or limiting magnitude is estimated based on the observed magnitude of the tracked star on the FPSC, which is then corrected based on magnitude gain $\Delta M_{\rm {snr}}$ from the ability to track at reduced signal-to-noise; this is further corrected by estimated magnitude gain $\Delta M_{\rm {exp,rn}}$ or $\Delta M_{\rm {exp,dc}}$ from increased exposure times limited by read-noise or dark-current, respectively; a maximum exposure time $t_{\rm exp, max} = $ 0.1\,seconds and minimum SNR $a_{\rm snr, min} = $ 10 is determined from simulation based on effective 20\,milliarcsecond image stabilization resolution; observed magnitudes reference the GAIA DR2 catalog\cite{gaia_dr2}, which has a similar band-pass to the overall effective throughput shown in Figure \ref{fig:fpsc_efficiency}.}
\begin{ruledtabular}
\begin{tabular}{ccc|cccc}
\multicolumn{3}{c}{Flight Data}\vline & \multicolumn{4}{c}{Magnitude Estimation}\\
Observed Mag. & $t_{\rm exp}$ [ms] & $a_{\rm snr}$ & $\Delta M_{\rm exp,dc}$ & $\Delta M_{\rm exp,rn}$ & $\Delta M_{\rm snr}$ & Limiting Mag.\\ \hline
8.7 & $14$   & $79$ & $1.0$ & $2.1$ & $2.2$ & 12--13\\
8.9 & $19  $ & $70$ & $0.9$ & $1.8$ & $2.1$ & 12--13\\
8.7 & $11$   & $85$ & $1.2$ & $2.4$ & $2.3$ & 12--14\\
6.0 & $ 1.0$ & $80$ & $2.5$ & $5.0$ & $2.3$ & 11--13\\
6.6 & $ 3.0$ & $54$ & $1.9$ & $3.8$ & $1.8$ & 10--12
\end{tabular}
\end{ruledtabular}
\end{table*}

Overall, FPSC depth not only affects how well the FGS can effectively stabilize the telescope focal plane, but also directly determines the availability of targets on the sky, where sensitivity to dimmer guide stars (i.e. higher apparent magnitude) increases the likelihood that a given target will have a trackable star on one or both of the FPSCs.
As such, it is important to assess the limiting star magnitude that the \superbit science telescope can use for image stabilization.
This was not directly measured during the 2019 flight due to time constraints, but the limiting star magnitude can be estimated with knowledge of system performance with known guide stars on the FPSC.
Specifically, Table \ref{tab:fpsc_mags} shows a summary of identified stars on the FPSCs at a given exposure time and signal-to-noise ratio (SNR) during unique image stabilization pointings during the 2019 flight.
In terms of apparent magnitudes $M$, the potential gains from increased exposure time ($t_{\rm exp}$) and decreased SNR ($a_{\rm snr}$) can be estimated by
\begin{eqnarray}
\Delta M_{\rm exp,rn} &=& 2.5\log\left(\tfrac{t_{\rm exp,max}}{t_{\rm exp}}\right)\\
\Delta M_{\rm exp,dc} &=& \tfrac{2.5}{2}\log\left(\tfrac{t_{\rm exp,max}}{t_{\rm exp}}\right)\\
\Delta M_{\rm snr} &=& 2.5\log\left(\tfrac{a_{\rm snr}}{a_{\rm snr,min}}\right).
\end{eqnarray}
Here, $\Delta M_{\rm exp,rn}$ and $\Delta M_{\rm exp,dc}$ bound the magnitude gain from increased exposure time assuming either read-noise- or dark-current-limited exposures, respectively, where the latter yields reduced magnitude gain due to increased noise that scales with $\sqrt{t_{\rm exp}}$. 

The minimum SNR is determined from simulation based on the fidelity of the centroiding algorithm used on the FPSCs, which suggests that the required one-tenth pixel centroid accuracy ($\sim$20\,milliarcseconds) is adequately maintained with $a_{\rm snr,min} = 10$.
For maximum exposure time, FGS simulations indicate that tracking stability -- limited by rate gyroscope noise (0.002$^\circ/\sqrt{\mbox{hr}}$) -- is sufficiently constrained at the 20\,milliarcsecond level by FPSC centroid estimation at 10\,Hz, or $t_{\rm exp,max} = 0.1$\,s (see control architecture\cite{RomualdezThesis18}).
Applying this to the 2019 tracking runs, Table \ref{tab:fpsc_mags} shows the limiting apparent magnitudes per target, with a flight average limiting magnitude range of of 11--13.
Surveying the GAIA DR2 star catalog\cite{gaia_dr2} a limiting magnitude $\leq$\,13, the FPSCs are within 1\arcmin of a guide star over about 83\% of the sky. 
When considering multiple angles of observation over the course of a night, this approaches 100\% of the sky.

For more conservative estimates, future \superbit science flights -- such as a 30--50 day mid-latitude flight -- would undoubtedly benefit from higher target availability over the full sky.
As such, additional sensitivity could potentially be gained by improving the optical efficiency of the back-end optics or the FPSC CCD, where the former would be a trade-off for sensitivity in the shorter wavelength near-UV bands with alternative optical coatings.
Improved target availability could also be achieved by increasing the proportion of the available telescope focal plane to FPSCs either through increased area per FPSC or additional FPSCs distributed about the science CCD.

\subsection{Telescope Opto-mechanical \& Focus Stability}
\label{ss:opto_mech_stability}
\subsubsection{Static Pointing Drift}
\begin{figure*}[t]
\centering
\includegraphics[width=0.93\textwidth]{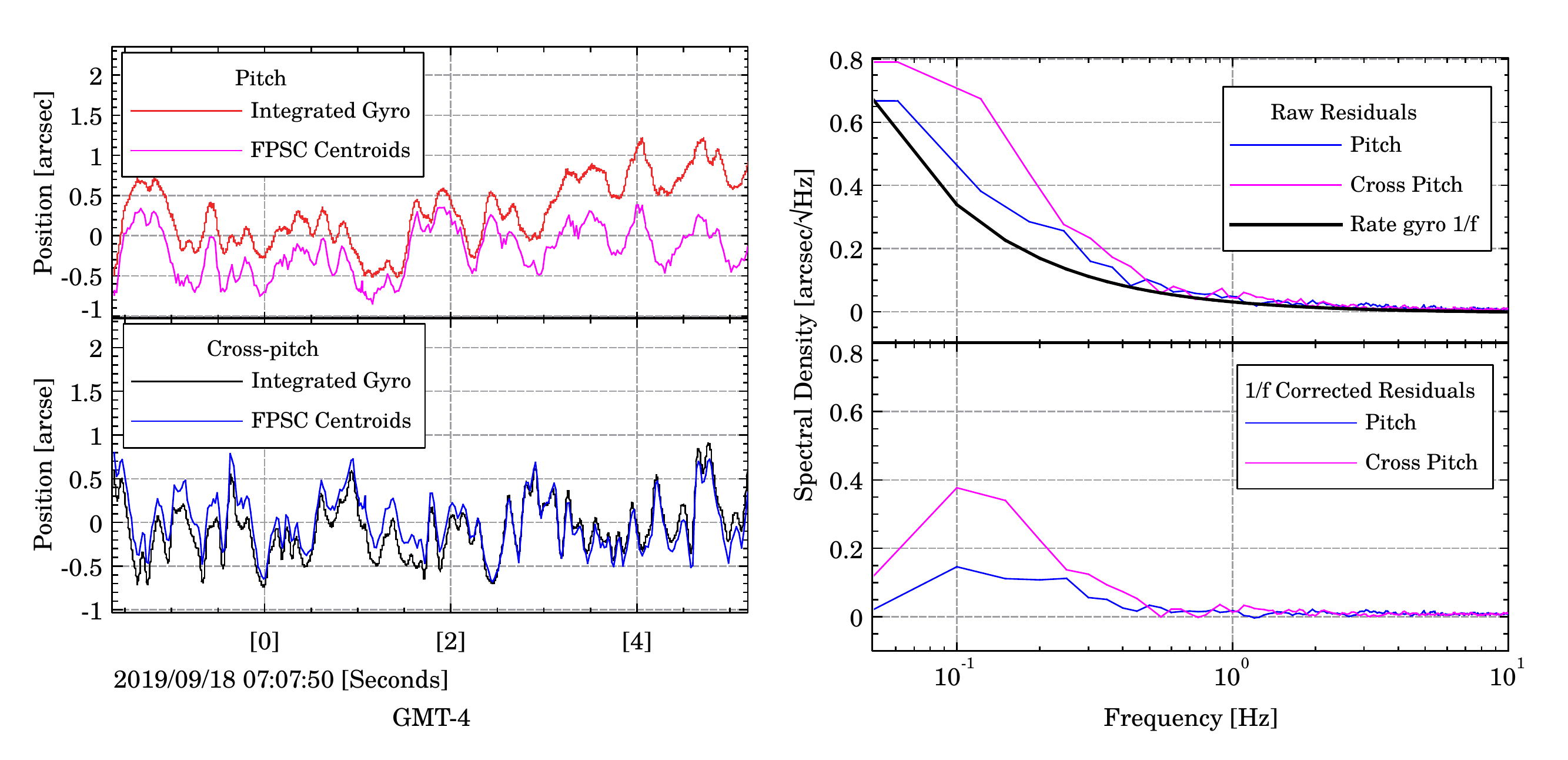}
\caption{Amplitude spectra of the computed difference between \superbit integrated rate gyroscopes and focal plane star camera centroids while the FGS tip-tilt stage is idle/not tracking (top right); gain differences and timing offsets between raw centroids and integrated rate gyroscopes are tuned manually with example timestreams compared in pitch (top left) and cross-pitch (bottom left); the 1/$f$ component from raw rate gyroscope integration (top right, solid black curve) is estimated and subtracted off (bottom right) with comparisons made up to the nominal control bandwidth of the telescope stabilization stage (10\,Hz); this shows relatively good agreement between integrated rate gyroscopes and focal plane centroids from 0.5--10\,Hz}
\label{fig:fpsc_gyro_centroids}
\end{figure*}
\superbit engineering flights prior to the 2019 flight had observed that relative pointing -- specifically between the boresight star camera and the main telescope -- had drifted significantly over long ($\geq$\,30\,minutes) tracking periods.
This effect had only been mitigated by the fact that the 10--15$\arcsec$ net sky-equivalent drifts observed had only been 18.7\% the full FGS throw in the pre-2019 configuration, where $\phi = 69.8\arcsec$ from (\ref{eq:fgs_throw}) with $f = 10$, $\ell = 168$ mm, $\theta = 10$\,mrad \cite{Romualdez16}. 

In contrast, the 2019 flight configuration performance results shown in Figure \ref{fig:pointing_drift} demonstrate much more stable relative pointing on the sky, with a worst case $\leq$\,1.0$\arcsec$ sky-equivalent drift over the same time period, or only 11.2\% the full cross-pitch throw of the FGS.
Note that compared to previous flight configurations, a 5 times reduction in FGS range -- as implemented for the 2019 flight -- favors improved position resolution over FGS throw.
Although this implies that the 2019 FGS would not have been able to compensate for the larger relative drifts observed pre-2019, the more rigidly and optically coupled mounting of the boresight star camera directly to the carbon fiber telescope baffle tube played a major role in reducing drift over long timescales by an order of magnitude.

In addition to this effect, apparent relative pointing drift can also be the product of relative motion between optical components within the telescope itself, namely the primary and secondary mirrors shifting within their respective mounts due to thermal changes or changes in gravitational loading at different elevations.
For pre-2019 \superbit test flights, the engineering telescope configuration utilized flexible, zero-stress mounting methods for optics (i.e. whiffle tree with spring side supports) typically used for ground-based or certain space-based applications. 
However, the large deflections observed from pre-2019 configurations may have also been indicative of significant gravity sag and, consequently, gross optical misalignment.
In contrast, the 2019 \superbit science telescope, as described in Subsection \ref{ss:superbit_architecture}, mitigates mechanical and thermal stress through static mounting of a conical primary, which is, by design, rigid to within the structural flexibility of the solid carbon fiber monocoque shell at the $\lambda$/20 surface roughness requirement.

\subsubsection{Relative Dynamics}
Although static boresight camera mounting drift cannot be easily decoupled from gravity sag of telescope and back-end optics, the relative dynamics between the telescope baffle (via rate gyroscopes) can be directly compared with FPSC centroids on the telescope focal plane.
As such, Figure \ref{fig:fpsc_gyro_centroids} shows the spectra of the residual pitch and cross-pitch differences between the FPSC centroids and raw integrated rate gyroscopes, which are rigidly coupled to the telescope tube assembly.
Since measurement took place during periods when the FGS was disabled despite having a trackable star on the FPSCs, this measurement represents the difference in frequency response between the telescope frame and the optics, assuming that external effects (e.g. stratospheric seeing at $<$\,0.010$\arcsec$) are negligible.

From pre-2019 \superbit test flights, significant disagreement had been observed at the 0.2$\arcsec$ level from 1--15\,Hz, where higher frequency motion had been captured by FPSC centroids that was not reflected in the rate gyroscopes.
To some level, this would imply dynamic instability had been present between the telescope frame and telescope optics -- likely due to the flexible primary mirror mount -- that could have been driven by the telescope stabilization stage.
In contrast, the 2019 flight results in Figure \ref{fig:fpsc_gyro_centroids} show significantly improved agreement between FPSC centroids and integrated rate gyroscopes, and therefore suggest better optical and dynamic stability between the science telescope optics and baffle frame.
Keeping in mind that $1/f$ drift is present in the raw integrated rate gyroscope timestreams, the sub-pixel agreement in amplitude spectra between integrated rate gyroscopes and FPSC centroids is further emphasized by the similar shape and magnitude of variations in the timestreams from 0.5--10\,Hz.
There does appear to be to be a slight residual below 0.5 Hz, but this can likely be attributed to spike removal in the timestreams (Emcore).

When considering the static drift (Figure \ref{fig:pointing_drift}), the relative dynamic stability (Figure \ref{fig:fpsc_gyro_centroids}), and the overall image stabilization performance (Figure \ref{fig:image_stability}), it is reasonable to assert -- in contrast with pre-2019 engineering test flights -- that the \superbit science telescope met mechanical stability specifications required to provide effective sub-pixel image stabilization at a level sufficient for diffraction-limited imaging.
Improvements in the rigidity of components' relative mounting or the fidelity of inertial measurements could potentially be informed by direct measurement of telescope optics motion during tracking (e.g. accelerometer measurements).

\label{s:summary}
\begin{table*}[t]
\caption{\label{tab:superbit_performance_todate}Summary of best achieved 3-axis absolute pointing and image stabilization performance for 4 \superbit test flights over 5\,years; stability over timescales representative of science camera integration periods (5\,minutes) are reported in addition to stability over extended periods (30\,minutes).}
\begin{ruledtabular}
\begin{tabular}{ccc|cc|cc}
\multicolumn{3}{c}{} \vline & \multicolumn{4}{c}{Best Achieved Sky-Fixed Stability ($1\sigma$) [arcseconds]}\\
\hline
Year & Launch Site & Provider & \multicolumn{2}{c}{Telescope Stabilization} \vline & \multicolumn{2}{c}{Image Stabilization}\\
 & & & @\,5\,min. & @\,30\,min. & @\,5\,min. & @\,30\,min.\\
 \hline
2015 & Timmins, ON & CNES-CSA & 0.5 & 1.5 & 0.085 & 0.5\\
2016 & Palestine, TX & CSBF-NASA & 0.5 & 1.1 & 0.070 & 0.2\\
2017/18 & Palestine, TX & CSBF-NASA & 0.4 & 0.8 & 0.065 & 0.090\\
2019 & Timmins, ON & CNES-CSA & 0.3 & 0.5 & 0.046 & 0.048

\end{tabular}
\end{ruledtabular}
\end{table*}

\subsubsection{FGS Depth-of-Focus Effects}
For \superbit's minimum wavelength in the near-UV ($\lambda = 300$\,nm), the minimum delta focus $\Delta F$ to induce a quarter-wavelength wavefront error is given by
\begin{eqnarray}
\Delta F \simeq \pm 2\lambda f^2
\end{eqnarray}
where $f$ is the $f$-number for the telescope.
For the pre-2019 engineering telescope with a 55\,mm diameter usable focal plane and $f=10$, the $\theta = 10$\,mrad full FGS throw could have potentially caused a defocus of 0.275\,mm at the edge, which exceeds the $\Delta F = 0.060$\,mm by nearly a factor of 5 had the full range been exercised.
However, the maximum drift experienced from previous test flights was only 18.7\% the full FGS throw equivalent to a maximum defocus of 0.514\,mm, which had been within the tolerable $\Delta F$ albeit marginally.

In contrast, the 2019 science telescope and back-end optics configuration has a five-fold reduction in FGS throw ($\theta = 2$\,mrad) with a maximum defocus of 0.055\,mm at the focal plane edge, which is comfortably within the $\Delta F = 0.0726$\,mm tolerable delta focus for $f=11$.
For depth-of-focus, this implies significant margin for the maximum observed $1\arcsec$ absolute drift observed during the 2019 flight (Figure \ref{fig:pointing_drift}), with nearly an order of magnitude more FGS throw than required to compensate.

For the 2019 science configuration overall, this ability to exercise the full FGS range highlights the flexibility in target re-acquisition accuracy during, for example, gimbal resets as shown in Figure \ref{fig:gimbal_reset}.
Furthermore, this potentially enables the trading-off of telescope stabilization gains for improved image stabilization, which may require the FGS to compensate for lower frequencies at high amplitude, as described previously.
Should additional range during image stabilization be required, as was likely the case for the pre-2019 \superbit configuration and could be the case for future \superbit flights for image dithering operations, FPSC centroid information could potentially be fed back to the telescope stabilization loop to mitigate larger low frequency perturbations.
However, care would have be to taken to ensure sufficient decoupling with image stabilization (i.e. prevent co-servoed jack-knifing).

\section{Summary \& Forecasting}
\label{s:forecast}
Table \ref{tab:superbit_performance_todate} shows the progression of \superbit performance over 4 test flights from 2015--2019 compared to the most recent science telescope commissioning flight in September 2019.
As previously mentioned, a major factor contributing to the improved performance of the \superbit 2019 flight compared with previous test flights was the design, implementation, and flight verification of the \superbit diffraction-limited telescope (Figure \ref{fig:telescope}), which provided the necessary opto-mechanical static and dynamic stability as well as optical beam quality (Figure \ref{fig:fpsc_coadd}) required for sufficiently robust focal plane stability.
This is clearly reflected in the image stabilization results for 2019 (Figure \ref{fig:image_stability}), where the performance over 5\,minute timescales is maintained over entire telescope tracking runs (Figure \ref{fig:telescope_stability}) at the 30--60\,minutes timescale. 
Improvements in 2019 opto-mechanical design and relative mounting stiffness allowed for higher resolution in the image stabilization stage over a smaller range compared to engineering test flights (Subsection \ref{ss:opto_mech_stability}).

From the latest 2019 performance, the level of image stability achieved has been shown to be sufficient for diffraction-limited imaging from over \superbit's wavelength range (300--1000\,nm) (Subsection \ref{ss:diffraction_limited}). 
As discussed in Subsection \ref{ss:stabilization}), 2019 flight performance could have been improved with more time allocated to trading-off the coarse stabilization loop performance at lower frequencies for reduced high frequency rate gyroscope noise that perturbed image stabilization (Figure \ref{fig:pitch_overcontrol}) through a careful gain reduction.
For the prospective \superbit mid-latitude LDB flight, this is effectively mitigated by the availability of additional calibration time on the sky as well as more opportunity for iteration with feedback from science images, both of which was quite limited by less than 8\,hours of operational time at float in 2019.

Ultimately, the optical and mechanical performance of the \superbit science telescope and gondola will determine effective science yield, specifically the number of clusters observed in the case of weak lensing.
The achievable depth on the science camera focal plane based on the telescope throughput will directly influence overall survey efficiency source completeness, while the PSF contributions from pointing jitter, optical alignment, and non-linear focal plane effects will impact the fidelity of background galaxy ellipticities at the 1--3\% level (estimated).
Although certain aspects of these science-related factors are explored in the work presented here, further work is currently being undertaken to fully assess \superbit weak lensing potential as well as other science forecasting.
This ongoing and future work will be captured in an upcoming \textit{\superbit forecasting paper}, in which the results of this work will be directly leveraged. 

\begin{acknowledgments}
Support for the development of SuperBIT is provided by NASA through APRA grant NNX16AF65G. 
Launch and operational support for the sequence of test flights from Palestine, Texas are provided by the Columbia Scientific Balloon Facility (CSBF) under contract from NASA's Balloon Program Office (BPO).
Launch and operational support for test flights from Timmins, Ontario are provided by the \textit{Centre National d'\'Etudes Spatiales} (CNES) and the Canadian Space Agency (CSA).

LJR is supported by the National Science and Engineering Research Council Post-doctoral Fellowship [NSERC PDF--532579--2019].
MJ is supported by the United Kingdom Research and Innovation (UKRI) Future Leaders Fellowship 'Using Cosmic Beasts to uncover the Nature of Dark Matter' [grant MR/S017216/1].

The Dunlap Institute is funded through an endowment established by the David Dunlap family and the University of Toronto.
UK coauthors acknowledge funding from the Durham University Astronomy Projects Award, the Van Mildert College Trust, STFC [grant ST/P000541/1], the Royal Society [grants UF150687 and RGF/EA/180026], and UKRI [grant MR/S017216/1].

This work has made use of data from the European Space Agency (ESA) mission {\it Gaia} (\url{https://www.cosmos.esa.int/gaia}), processed by the {\it Gaia} Data Processing and Analysis Consortium (DPAC, \url{https://www.cosmos.esa.int/web/gaia/dpac/consortium}). 
Funding for the DPAC has been provided by national institutions, in particular the institutions participating in the {\it Gaia} Multilateral Agreement.
Additionally, this work made use of the \texttt{SAOImage DS9} imaging application \cite{ds9}, \texttt{Astrometry.net} \cite{astrometry}, and \texttt{SExtractor} \cite{sextractor}.
\end{acknowledgments}

\bibliographystyle{aipnum4-1}
\bibliography{references}

\end{document}